\documentclass[aps,prd,twocolumn,showpacs,showkeys,amsmath,amssymb]{revtex4}
\usepackage{graphicx}% Include figure files
\usepackage{epstopdf}
\usepackage{dcolumn}% Align table columns on decimal point
\usepackage{bm}% bold math
\usepackage{amssymb}
\usepackage{amsmath}
\usepackage{color}
\usepackage[serbian,english]{babel}

\begin{document}

\title{Angular momentum and topological dependence of Kepler's Third Law in 
the Broucke-Hadjidemetriou-H\' enon family of periodic three-body orbits} 
\author{Marija R. Jankovi\' c}
\affiliation{Faculty of Physics, %Fizi\v cki fakultet, 
Belgrade University,\\
Studentski Trg 12, 11000 Belgrade, Serbia}
\author{V. Dmitra\v sinovi\' c}
\affiliation{Institute of Physics, Belgrade
University, Pregrevica 118, Zemun, \\
P.O.Box 57, 11080 Belgrade, Serbia} 
%marija.jankovic91@ymail.com
%, Milovan \v Suvakov$^{*}$}
%{\it Institute of Physics, Belgrade University, Pregrevica 118,
%Zemun,} \\
%{\it P.O.Box 57, 11080 Beograd, Serbia}}
% \\{\rm * currently on leave of absence at} Serbian Ministry of Education, Science and
%Technological Development, Nemanjina 22, 11050 Beograd, Serbia }

\date{\today}

\begin{abstract}
We use 57 recently found topological satellites of Broucke-Hadjidemetriou-H\' enon's periodic orbits 
with values of the topological exponent $k$ ranging from $k$ = 3 to $k$ = 58 to plot the 
angular momentum $L$ as a function of the period $T$, with both $L$ and $T$ rescaled to energy 
$E=-\frac12$.  Upon plotting $L(T/k)$ we find that all our solutions fall on a curve 
that is virtually indiscernible by naked eye from the $L(T)$ curve for non-satellite solutions. 
The standard deviation of the satellite data from the sixth-order polynomial fit
to the progenitor data is $\sigma = 0.13$. %We point out that t
This regularity 
supports H\' enon's 1976 conjecture that the linearly stable Broucke-Hadjidemetriou-H\' enon orbits are also perpetually,
or Kol’mogorov-Arnol’d-Moser stable.
\end{abstract}

\pacs{45.50.Jf, 05.45.-a, 95.10.Ce}

\keywords{celestial mechanics; three-body systems in Newtonian gravity; nonlinear dynamics} 

\maketitle

\section*{Introduction}

Numerical studies of periodic three-body orbits have increased their output over the past few years - 
more than 40 new orbits, and their ``satellites'' have been discovered, Refs. \cite{Suvakov:2013,Simo2002,Suvakov:2013b,Shibayama:2015}.
Unlike periodic two-body orbits, which are all ellipses, and thus are all topologically equivalent, %to a simple loop, %circle, 
the non-colliding three-body periodic orbits have one of infinitely many different topologies.
Montgomery, Ref. \cite{Montgomery1998}, had devised an algebraic method to associate a 
free-group element (``word'') $w$ with a three-body orbit's topology, and thus to label and 
classify such periodic orbits; for an elementary introduction to this method, see 
Ref. \cite{Suvakov:2014}. That classification method 
% was merely a subject of academic discussions until {\color{red} 
has recently acquired practical importance in the identification of new three-body 
orbits, Refs. \cite{Suvakov:2013,Suvakov:2013b,Shibayama:2015}.

A number of newly discovered orbits, Refs. \cite{Suvakov:2013,Simo2002,Suvakov:2013b,Shibayama:2015},  
were of the so-called topological satellite type. Such satellite orbits, 
are also known as ``bifurcation'' in the older literature, Refs. \cite{Simo2002,Davoust1982}, 
where they were only loosely defined in terms of their presumed origin. 
It was only in Ref. \cite{Suvakov:2013b} that a precise definition of a 
topological satellite was given.
When this definition was applied to the figure-8 satellites 
\footnote{Satellite orbits of the figure-eight were first observed in Ref. \cite{Simo2002} 
and further investigated in Refs. \cite{Suvakov:2013b,Shibayama:2015}}, 
reported in Ref. \cite{Suvakov:2013b}, it led to the discovery of a remarkable 
``topological Kepler's third law''-like regularity  
for arbitrary orbits with vanishing angular momenta, Ref. \cite{Dmitrasinovic:2015}. 
The immediate question is whether this regularity persists when the angular momentum 
does not vanish? 

The present Letter is an attempt to answer that question, albeit in a single, specific family 
of three-body orbits, {\it viz.} in the Broucke-Hadjidemetriou-H\' enon (BHH) family 
\cite{Broucke1975,Broucke1975b,Hadjidemetriou1975a,Hadjidemetriou1975b,Hadjidemetriou1975c,Henon1976,Henon1977}, 
that has the simplest non-trivial topology (free group element $w$={\tt a}). 
The main reason for selecting only this family of orbits is that it is the most thoroughly 
studied family thus far: it is the only family of orbits with a previously determined dependence 
of the period $T$ on the angular momentum $L$ of (non-satellite, or progenitor) periodic orbits, 
Refs. \cite{Broucke1975,Broucke1975b,Hadjidemetriou1975a,Hadjidemetriou1975b,Hadjidemetriou1975c,Henon1976,Henon1977}. 
No such, or comparable, study of any of the remaining known families exists to our knowledge at 
this moment. 
%{\color{red} 
Moreover, the BHH family is one of only two families \footnote{The other one being 
the Lagrange family of orbits.} of periodic three-body orbits that have been observed in 
astronomy: all known ``hierarchical'' triple star systems belong to BHH orbits.
Moreover, the Sun-Earth-Moon system may be viewed as a BHH solution, albeit with highly
asymmetrical mass ratios.

The first step towards this goal, the one of finding as many different BHH satellite orbits as possible, 
has already been accomplished in Ref. \cite{Jankovic:2013}. Previously, Davoust and 
Broucke, Ref. \cite{Davoust1982}, had found one (the first $k$=3) satellite of one retrograde 
BHH orbit. 
Ref. \cite{Jankovic:2013} extended the search for retrograde BHH satellite orbits systematically 
up to values $k \leq 19$ of the topological exponent $k$, and more haphazardly up to $k=58$; 
thus several different types of BHH satellites with identical values of $k$ were discovered,
\footnote{The presence of multiple satellites with the same topology is not the first known 
instance of its kind: there are (many) different satellites of the figure-eight orbit with 
identical values of $k$, see Ref. \cite{Suvakov:2013b,Shibayama:2015}, albeit with zero angular 
momentum.}, as were a few prograde BHH satellites, see the Supplemental Material \cite{Supplement} 
and the Web site \cite{Gallery}. 
%As the satellite orbits represent different (local) minima of the action, the question naturally
%arises: just how many such minima are there for each value of $k$?}
Prograde BHH satellites have not been studied systematically, as yet, mostly due to their paucity
at the values of the angular momentum covered in the searches in Ref. \cite{Jankovic:2013}.
%{\color{red} 
Presently it is not known how many satellites ought to exist, and under which 
conditions. It is interesting, however, that the observed satellites correspond only 
to linearly stable BHH progenitor orbits. 
This is in line with H\'enon's 1976 conjecture \cite{Henon1976,Supplement} about Kol’mogorov-Arnol’d-Moser (KAM) stability of linearly 
stable BHH orbits.

Then, motivated by the findings reported in Ref. \cite{Dmitrasinovic:2015}, we checked 
for similar regularities of satellite BHH orbits with non-zero angular momentum. 
Firstly, we formulated the topological dependence of Kepler's third
law for three-body orbits with non-zero angular momenta, and secondly we tested it on
the presently known satellites of the retrograde BHH family. %{\color{red} 
Secondly we found 
a striking result: all of our retrograde BHH satellites fall on a single (continuous) curve 
$L(T/k)$, Fig. \ref{fig:sat_TL_interpolated} that is practically indiscernible by naked eye 
from the $L(T)$ curve, Fig. \ref{fig:LT_BHH}, for non-satellite (progenitor) retrograde BHH 
solutions, whereas the ``topologically uncorrected'' curve $L(T)$ looks very differently, 
see Fig. \ref{fig:sat_retro_TL}. 
A quantitative measure of this (dis)agreement is shown in terms of 
corresponding standard deviations.

\section*{Preliminaries}
\label{s:Preliminaries}

%\subsection*{Basics}
%\label{ss:Basics}

\begin{figure}
\includegraphics[width=0.45\textwidth]{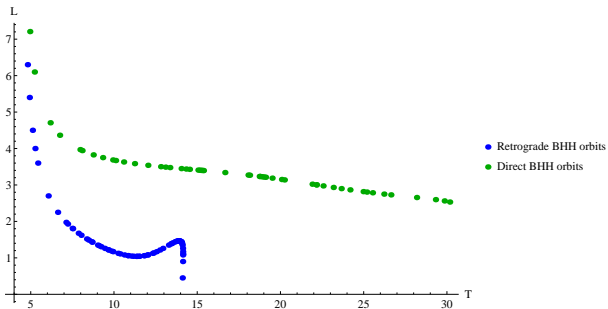}
%{E=0.5.jpg}%,\includegraphics[width=0.45\textwidth]{BHHretrograde.eps}%{T-L_sve_orbite.eps}
\caption{$L(T)$ curves for direct, or prograde (green, upper set of points) and 
retrograde (blue, lower set of points) BHH orbits, all at fixed energy $E=-0.5$.} 
\label{fig:LT_BHH} 
\end{figure}

Broucke \cite{Broucke1975,Broucke1975b,Davoust1982}, Hadjidemetriou 
\cite{Hadjidemetriou1975a,Hadjidemetriou1975b,Hadjidemetriou1975c} 
and H\' enon \cite{Henon1976,Henon1977} (BHH) explored a set of periodic planar three-body 
orbits with equal mass bodies. These orbits form two continuous curves in the L-T plane 
%relative periodic orbits in the phase space of initial conditions, 
whose lower (retrograde) terminus (``end'') is the collinear collision (Schubart) orbit, 
and both the retrograde and the direct $L(T)$ curves approach the same high-L limit 
at their upper termini, Fig. \ref{fig:LT_BHH}.
%\onecolumngrid

Although BHH write of two families of orbits - direct, or prograde, and retrograde - all of 
these orbits belong to a single topological family: during one period the orbit completes a 
single loop around one of the poles on the shape sphere. 
This loop can be described by the conjugacy class of the fundamental group/free group element {\tt a}, 
according to the topological classification used in Refs. \cite{Suvakov:2013,Suvakov:2014}.
It turns out, however, that there are numerous relative periodic orbits with topology 
${\tt a}^k$, with $k=2,3..$. % that have the same form of initial conditions. 
Such orbits are sometimes called satellites \cite{Simo2002,Suvakov:2013b}, whereas other 
authors call them ``bifurcation orbits'' \cite{Davoust1982}. 

\subsection*{Scaling laws for three bodies}
\label{ss:Scaling}

It is well known that Kepler's third law (for two bodies) follows from the spatio-temporal 
scaling laws, which, in turn, follow from the homogeneity of the Newtonian gravity's static 
potential, Ref. \cite{Landau}. These scaling laws read 
${\bf r} \rightarrow \lambda {\bf r}$, $t \rightarrow \lambda^{3/2} t$, and 
consequently ${\bf v} \rightarrow {\bf v} / \sqrt{\lambda}$. The (total) energy
scales as $E \rightarrow \lambda^{-1}E$, the period $T$ as 
$T \rightarrow \lambda^{3/2} T$ and angular momentum as $L \rightarrow \lambda^{1/2} L$, 
i.e., differently than either the period $T$, or ``size'' $R$, which is the reason why 
only the vanishing angular momentum $L=0$ is a ``fixed point'' under scaling.
For this reason, we use scale-invariant angular momentum $L_r = L |E|^{1/2}$, 
scale-invariant period $T_r = T |E|^{3/2}$ and, for simplicity's sake, equal masses. 
Thus, we may replace the ``mean size'' ${\bar R}$ of the three-body system in Kepler's third law 
$T \propto {\bar R}^{3/2}$ with the inverse absolute value of energy $|E|^{-1}$, i.e., 
$T \propto |E|^{-3/2}$, or equivalently $T|E|^{3/2} = T_r = {\rm const.}~$. 

The ``constant'' on the right-hand-side of this equation %$T|E|^{3/2} = {\rm const.}~$
is not a universal one in the three-body case, as it is in the two-body case (where it 
depends only on the masses and the Newtonian coupling $G$): 
it may depend both on the family $w$ of the three-body orbit, described by the free-group word $w$, 
%described by $T(w^k)|E(w^k)|^{3/2} = k T(w)|E(w)|^{3/2}$, or equivalently $T_r (w^{k}) = k T_r(w)$, 
%for orbits with non-zero angular momenta, see Ref. \cite{Dmitrasinovic:2015}, 
and on the scale-invariant angular momentum $L_r = L |E|^{1/2}$ of the orbit, 
see Refs. \cite{Henon1976,Henon1977}, as follows 
\[ T^{(w)}|E|^{3/2} = T_r^{(w)} = f(L^{(w)} |E|^{1/2}) = f(L_r^{(w)}),\]
or as an inverse function:
\[ L_r^{(w)} = L^{(w)}|E|^{1/2} = f^{-1}(T^{(w)} |E|^{3/2}) =  f^{-1}(T_r^{(w)}).\]
Thus, the curve $L_r^{(w)}(T_r^{(w)})= L^{(w)}|E|^{1/2}(T^{(w)}|E|^{3/2})$ as a function of 
$T_r^{(w)} = T^{(w)}|E|^{3/2}$ is a fundamental property of any family $w$ of periodic orbits.
For the BHH family the L(T) curve, for fixed energy $E=-0.5$ orbits, based on the data from Refs. 
\cite{Broucke1975,Broucke1975b,Hadjidemetriou1975a,Hadjidemetriou1975b,Hadjidemetriou1975c,Henon1976,Henon1977} 
is shown in Fig. \ref{fig:LT_BHH}.

We wish to see if the zero-angular-momentum relation $T_r (w^{k}) = k T_r(w)$, Ref. \cite{Dmitrasinovic:2015}, 
or some similar statement holds also at non-zero angular momentum? The analogon of this relation 
for orbits with non-zero angular momenta would be a simple relation between L(T) curves for the progenitor
orbit $L_r(T_r)$ and its $k$-th satellite $L_r^{(w^k)}(T_r^{(w^k)})$: 
\begin{eqnarray} 
L_r^{(w)}(T_r^{(w)}) &=& L_r^{(w^k)}(T_r^{(w^k)}/k).
\end{eqnarray}
We shall test this relation in the BHH family of solutions, and in order to do so, 
we use %need %must first find to have
the BHH satellite orbits from Ref. \cite{Jankovic:2013}.

\subsection*{L(T) curves for BHH satellites}
\label{ss:LT_curves}

\begin{figure}
\includegraphics[width=0.45\textwidth]{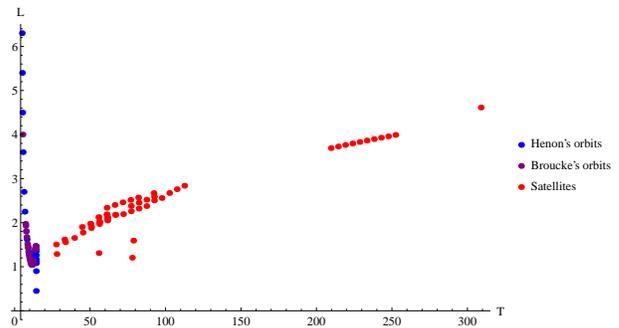}
\caption{L(T) dependence of retrograde BHH orbits' (blue dots of different hues) 
and their satellites' (red), with various values of $k$, all at fixed energy $E=-0.5$. The 
data are from Table \ref{tab:overview1_norm}.}
\label{fig:sat_retro_TL} 
\end{figure}

\begin{figure}
\includegraphics[width=0.45\textwidth]{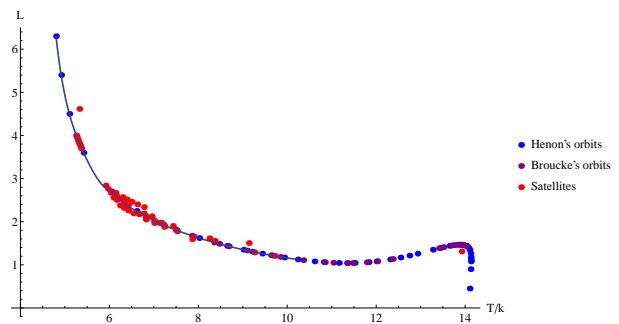}
\caption{L(T'=T/k) dependence at fixed energy $E=-0.5$ for the aggregate set of retrograde 
BHH orbits (blue dots of different hues) and their satellites (red dots) 
with various values of $k$,
together with the fitted interpolating function (blue solid). The data are from 
Table \ref{tab:overview1_norm}.}
\label{fig:sat_TL_interpolated} 
\end{figure}

\begin{figure}
\includegraphics[width=0.45\textwidth]{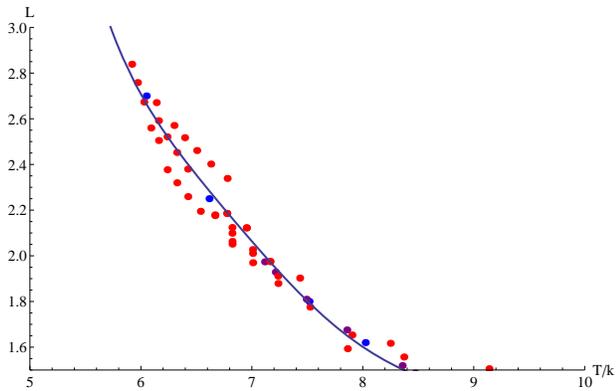}
\caption{Enlargement of the L$\in [1.5,3]$ region of the retrograde BHH orbits (blue dots) 
and their satellites (red dots) with various values of $k$ L(T'=T/k) dependence at fixed energy $E=-0.5$.
Note that the size of the dots on the diagram exceeds the corresponding numerical uncertainties 
(``error bars'').}
\label{fig:sat_TL} 
\end{figure}

The L-T plot of different-$k$ satellite orbits are scattered over a large region and do not 
intersect the BHH progenitor family of orbits' L(T) curve when plotted as a function 
of the (un-divided) period T, see Fig. \ref{fig:sat_retro_TL}. Note the large span  
of periods T in the data, Table \ref{tab:overview1_norm}, and in Fig. \ref{fig:sat_retro_TL}, 
as well as two large ``gaps'' in the data. These gaps are due to the exigencies of the search 
reported in Ref. \cite{Jankovic:2013}, which was not conducted with 
the intention of testing the hypothetical topological Kepler's third law.
%===============================================================
\begin{table}[h!]
\begin{center} 
\caption{Properties of satellite orbits in the retrograde branch of the BHH family. 
% shown in %Fig. \ref{fig:overview}, \ref{fig:satellites_small_k}. 
Here $k$ is the topological power of the orbit, T is its period, and L its angular momentum. 
All orbits have the same energy $E=-\frac12$. 
For the raw data and a discussion of numerical errors, 
%and the minimal return proximity for these orbits is $d_{\rm min}<10^{-3}$, 
see the Supplemental Material \cite{Supplement}.}
\begin{tabular}{c@{\hskip 0.2in}c@{\hskip 0.2in}c@{\hskip 0.1in}c@{\hskip 0.2in}
c@{\hskip 0.2in}c@{\hskip 0.1in}}
\hline \hline 
\setlength
%$N_r$ & 
$T$ & $L$ & ${k}$ %& %$N_r$ 
& $T$ & $L$ & ${k}$ \\
\hline
\hline
%1	&	{\color{red}*}
27.80080	&	1.28815	&	3	&	71.53838	&	2.46095	&	11	\\
27.41157	&	1.50552	&	3	&	77.07474	&	2.25918	&	12	\\
32.99245	&	1.61682	&	4	&	77.06060	&	2.37981	&	12	\\
33.47935	&	1.55701	&	4	&	76.73111	&	2.51718	&	12	\\
55.67884	&	1.31000	&	4	&	82.21327	&	2.31968	&	13	\\
39.51102	&	1.65331	&	5	&	82.19918	&	2.45231	&	13	\\
45.13827	&	1.77568	&	6	&	81.88258	&	2.57068	&	13	\\
44.58632	&	1.90240	&	6	&	87.31760	&	2.37687	&	14	\\
50.64660	&	1.87900	&	7	&	87.30360	&	2.52098	&	14	\\
50.63890	&	1.91139	&	7	&	92.38479	&	2.50486	&	15	\\
50.14113	&	1.97452	&	7	&	92.37738	&	2.59166	&	15	\\
50.14128	&	1.97537	&	7	&	92.08210	&	2.67070	&	15	\\
56.06083	&	1.96971	&	8	&	97.43210	&	2.55979	&	16	\\
55.60411	&	2.12189	&	8	&	102.45058	&	2.67331	&	17	\\
77.81366	&	1.20544	&	8	&	107.44964	&	2.75861	&	18	\\
56.05269	&	2.01054	&	8	&	112.42918	&	2.83883	&	19	\\
56.04953	&	2.02709	&	8	&	209.48795	&	3.69220	&	39	\\
55.60430	&	2.12289	&	8	&	214.25815	&	3.72785	&	40	\\
61.39903	&	2.05128	&	9	&	219.02302	&	3.76283	&	41	\\
60.96889	&	2.18581	&	9	&	223.78278	&	3.79719	&	42	\\
61.39086	&	2.09890	&	9	&	228.53763	&	3.83094	&	43	\\
61.38676	&	2.12397	&	9	&	233.28775	&	3.86412	&	44	\\
60.96879	&	2.18532	&	9	&	238.03332	&	3.89675	&	45	\\
60.99996	&	2.33882	&	9	&	242.77450	&	3.92885	&	46	\\
61.39697	&	2.06300	&	9	&	247.51146	&	3.96044	&	47	\\
66.66644	&	2.17917	&	10	&	252.24433	&	3.99155	&	48	\\
66.66689	&	2.17608	&	10	&	308.85330	&	4.61404	&	58	\\
66.29761	&	2.40165	&	10	&	& &					\\
78.61058	&	1.59325	&	10	&	& &					\\
71.89715	&	2.19481	&	11	&	& &					\\
\hline
\end{tabular}
\label{tab:overview1_norm}
\end{center}
\end{table}
%===============================================================
The values in Table \ref{tab:overview1_norm} have been rounded off to five 
significant decimal places. So, the numerical error is less than one part in 10,000. 
Such an error would be invisible in the Figs. \ref{fig:sat_retro_TL},\ref{fig:sat_TL_interpolated},\ref{fig:sat_TL} 
meaning that the ``size of the points'' in these figures is larger than the expected error.
%\onecolumngrid
%\twocolumngrid
After dividing the period T (at fixed energy) by the topological exponent %/index 
$k$, ${\rm T}^{'} = {\rm T}/k$, 
we can see in Fig. \ref{fig:sat_TL_interpolated} that the satellite orbits' L(T/$k$) curve (the angular 
momentum L as a function of topologically-rescaled period T/$k$) approximately coincides with 
the L(T) curve of BHH retrograde orbits. 
It seems that such an appearance of order out of apparent disorder cannot be an accident.

Next, in Fig. \ref{fig:sat_TL_interpolated} we look more closely at the section of the L(T) curve of 
progenitor BHH retrograde orbits in which we have found all but one of our satellites. 
We have interpolated H\' enon's, \cite{Henon1976}, 
%(how many? 18 H\' enon's plus 10 Broucke's = 28) 
18 stable retrograde data points with a piecewise polynomial fit in this part of the L(T) curve.
The standard deviations from this interpolated curve were calculated for: 1) Broucke's 10 
progenitor retrograde orbits, \cite{Broucke1975,Broucke1975b}, and 
2) the 56 out of 57 new satellite orbits from Table %\ref{tab:preliminary},
\ref{tab:overview1_norm} (excluding one orbit that lies near the ``shoulder'' at T=14 in Fig. \ref{fig:sat_TL_interpolated}), 
with the following results. 1) $\sigma=0.0034$ for Broucke's orbits; and 2) $\sigma=0.1269$ for 
satellite orbits. 
This difference of two orders of magnitude between these two numbers clearly indicates 
that the rescaled satellites' periods do {\it not} coincide {\it exactly} with the progenitor 
ones, but only approximately. 

Moreover, when one assembles H\' enon's and Broucke's, \cite{Broucke1975,Broucke1975b}, retrograde orbits 
in one set and fits the aggregate data by a polynomial of the sixth degree, Fig. \ref{fig:sat_TL_interpolated}, 
%then 
the standard deviation of the fit is 
$\sigma=0.0313$, %0.0312792$ 
whereas the standard deviation of all satellite orbits from this polynomial curve is 
$\sigma=0.1315$, %0.131474
roughly four times bigger. It is (statistically) clear that the satellites do not follow exactly the 
same L(T) curve as the progenitors, but the deviation is not large.
This constitutes the evidence for the analogon of the topological dependence of Kepler's third 
law for the $L \neq 0$ case, Ref. \cite{Dmitrasinovic:2015}.

Finally, we note that all of our newly found satellite orbits fall into a region of the progenitor 
L(T) curve that corresponds to stable progenitor BHH orbits, with one possible exception (the red point 
near the ``shoulder'' at T=14 in Fig. \ref{fig:sat_TL_interpolated}), that ``sits'' on the border point 
between stable and unstable regions. We have not found any other satellites in this, the second stable  
region of BHH retrograde orbits. In Fig. \ref{fig:sat_TL} we show the fine structure in the 
satellites' L(T/$k$) curve, that remains to be studied in finer detail and be better understood.

We have not studied the direct/prograde (sub)family of BHH orbits, as Ref. \cite{Jankovic:2013} 
did not search for their satellites, but found four %of them  
almost inadvertently. Certainly, that task ought to be completed in the future.

\section*{Summary, Conclusions and Outlook}
\label{s:Summary}

We have used %more than 
57 new satellite orbits from Ref.  \cite{Jankovic:2013}, %have been numerically found, 
in the family of Broucke-Hadjidemetriou-H\' enon, Refs. 
\cite{Broucke1975,Broucke1975b,Henon1976,Henon1977,Hadjidemetriou1975a,Hadjidemetriou1975b,Hadjidemetriou1975c}, 
relative periodic solutions to the planar three body problem.
Thence followed a striking relation between their kinematic and topological properties. 

BHH orbits constitute a family with a simple topology, described by the free group element {\tt a} 
according to the classification on the shape sphere, and their satellites are orbits of the topology 
${\tt a}^k$. The BHH orbits' angular momenta L and periods T form a continuous curve L(T), at fixed energy. 
Our satellite orbits form a scattered set of points on the same L(T) plot, but all of them exhibit the 
property that after their period T is divided by their topological
order $k$, they approximately fall on the L(T) curve of the original ($k=1$) BHH orbits.

This study was motivated by the discovery, Ref. \cite{Dmitrasinovic:2015}, of a relation between 
the topology and periods among the satellites of the figure-eight orbit, Ref.\cite{Suvakov:2013b}, 
and one other type (``moth I'' - ``yarn'' in Ref. \cite{Suvakov:2013}), of three-body orbits at 
vanishing angular momentum. 
%for which the same property was found. 
%Further Kepler-like topological regularities were found not just in 
%the satellite-progenitor relation, Ref. \cite{Dmitrasinovic:2015}, all at vanishing angular momenta.
%It was believed that such a relation would hold only at zero angular momentum $L = 0$.
This Letter shows that Kepler's third law's topological dependence also holds for orbits with 
$L \neq 0$, albeit only approximately. It remains to be seen just precisely what this discrepancy 
depends on?

These results are even more striking if one remembers that among our results there are several 
distinct types of satellite orbits of the same topological power $k$, some with quite different 
values of L and T, which all display this property. 
A closer look at the L(T/$k$) curve revealed a fine structure, which
should be investigated in higher detail in the future.
An extension of the search conducted in Ref. \cite{Jankovic:2013}, into hitherto unexplored 
regions of the L-T plane ought to provide (new) data that will further test our result.

Our results indirectly confirm H\'enon's 1976 conjecture, see page 282 in 
Ref. \cite{Henon1976}, reproduced in the Supplemental Material \cite{Supplement}, 
that the linearly stable BHH orbits are also nonlinearly, or perpetually, or 
KAM stable. Such KAM stability implies the existence of 
quasiperiodic orbits with periods that conform to the quasiperiodicity condition 
(i.e. with periods that are ``almost commensurate'' with the BHH progenitor's period), 
as predicted by the KAM theorem, Refs. \cite{Kolmogorov:1954,Arnold:1963a,Moser:1962}.

Our study opens several new questions:
1) The most commonly observed hierarchical triple star systems belong to the BHH family.
%For this reason i
Are there BHH topological satellites among astronomically observed three-body systems?
It is important to extend the present study to the realistic case of three different 
masses: some early 
work has already been done in this direction by Broucke and Boggs, Ref. \cite{Broucke1975}, and 
by Hadjidemetriou and Christides, Ref. \cite{Hadjidemetriou1975b}.
2) In recent years there have been %progress in providing 
formal ``proofs of existence'' given for at least some BHH orbits, 
Refs. \cite{Chen2008,Chen2009}. This begs the question: can one 
``prove existence'' of their satellite orbits, and, if yes, of 
how many satellites, and under which conditions?

\section*{Acknowledgments}

M. R. J. was a recipient of the ``Prof Dr Djordje {\v Z}ivanovi{\' c}'' scholarship 
%for the academic year 2013/14, 
awarded jointly by the Faculty of Physics and 
the Institute of Physics, Belgrade University, and was also supported by a 
City of Belgrade studentship (Gradska stipendija grada Beograda).
The work of V. D. was supported by the Serbian Ministry of Science and
Technological Development under grant numbers OI 171037 and III 41011.
The computing cluster Zefram (zefram.ipb.ac.rs) at the Institute of Physics Belgrade 
has been extensively used for numerical calculations.

\end{document}